\documentclass[aps,prl,superscriptaddress,twocolumn]{revtex4-1}
\usepackage{graphicx}
\usepackage{amsmath}
\usepackage{mathtools}
\usepackage{amsfonts}
\usepackage{bm}
\usepackage[normalem]{ulem}
\usepackage[usenames, dvipsnames]{xcolor}
\usepackage[colorinlistoftodos]{todonotes}

\usepackage{hyperref}
\hypersetup{colorlinks=true, linkcolor=blue, citecolor=blue, urlcolor=blue}

\begin{document}

\title{\textit{Ab initio} many-body \textsl{GW} correlations in the electronic structure of LaNiO$_2$}

\author{Valerio Olevano}
\affiliation{CNRS, Institut N\'eel, 38042 Grenoble, France}
\affiliation{Universit\'e Grenoble Alpes (UGA), 38000 Grenoble, France}

\author{Fabio Bernardini}
\affiliation{Dipartimento di Fisica, Universit\`a di Cagliari, 09042 Monserrato, Italy}

\author{Xavier Blase}
\affiliation{CNRS, Institut N\'eel, 38042 Grenoble, France}
\affiliation{Universit\'e Grenoble Alpes (UGA), 38000 Grenoble, France}

\author{Andr\'es Cano}
\affiliation{CNRS, Institut N\'eel, 38042 Grenoble, France}
\affiliation{Universit\'e Grenoble Alpes (UGA), 38000 Grenoble, France}

\date{\today}

\begin{abstract}
We present an \textit{ab initio} $GW$ self-energy calculation of the electronic structure of LaNiO$_2$.
With respect to density-functional theory we find that in $GW$ the La~4$f$ states undergo an important $+$2 eV upward shift from the Fermi level, while the O~2$p$ states are pulled down by $-$1.5~eV, thus reinforcing the charge-transfer character of this material.
However, $GW$ many-body effects leave the $d$-like bands at the Fermi level almost unaffected, so that the Fermi-surface topology is preserved, unlike in cuprates.
\end{abstract}

\maketitle

\paragraph{Introduction.}
The recent discovery of superconductivity in Sr-doped NdNiO$_2$/SrTiO$_3$ thin films \cite{li-nature19} has attracted enormous interest
\cite{botana19,wu19,sakakibara19,bernardini19a,bernardini2019b,bernardini2020a,nomura19,hepting19,zhang19a,zhang19,jiang19,zhou19,hirsch19,hu19}.
Such a discovery could represent the first successful extension of superconductivity from cuprates to isostructural/isoelectronic nickelates.
Indeed the bulk nickelates NdNiO$_2$ and LaNiO$_2$ share the same crystal structure with the cuprate superconductor CaCuO$_2$, and present also a very similar band structure at the level of density-functional theory (DFT) \cite{anisimov99,pickett-prb04,botana19,wu19,sakakibara19,bernardini19a}.
Thus, if the discovery will be confirmed \cite{zhou2019absence,li2019absence}, these nickelates could provide an important workbench to check some key ideas put forward in relation to the microscopic mechanism of high-temperature unconventional superconductivity.
In this respect, the most popular paradigm relies on the physics of strong correlations as described by the Hubbard model which provides the anti-ferromagnetic (AFM) parent phase of cuprates as a Mott insulator \cite{sawatzky19}.
Thus, according to this picture, the emergence of superconductivity seems to require the presence of such an AFM Mott insulator state in first place.
This point would be severely questioned if superconductivity in nickelates and cuprates is confirmed to have a common microscopic origin.
Indeed nickelates are paramagnetic metals in their parent phase \cite{HaywardRosseinsky03,sawatzky19}.
Thus, it is important to carefully clarify the analogies and differences between cuprates and nickelates, and in particular to study their electronic structure.
On nickelates most of the studies have been carried out so far within the framework of density-functional theory in the local-density approximation (LDA) or beyond \cite{anisimov99,pickett-prb04,botana19,wu19,sakakibara19,nomura19,hepting19,bernardini2019b}.
To make further progress it is therefore important to establish how these features are affected by electronic correlations. 
Attempts in this direction were carried out within dynamical mean-field theory (DMFT) \cite{RyeeHan19,lechermann19,KarpMillis20}.

In this work we calculate the electronic structure of LaNiO$_2$ within the framework of \textit{ab initio} many-body perturbation theory taking into account correlations in the $GW$ approximation for the self-energy \cite{MartinReiningCeperley}.
We restrict ourselves to first-order perturbation theory one-iteration $G_0W_0$, starting from DFT-LDA as the unperturbed zero order.
We find that the $d$ bands around the Fermi surface are almost unaffected by many-body effects and that $GW$ preserves the DFT-LDA Fermi-surface topology, with only a moderate change in the size of the electron pockets.
The Ni~3$d_{x^2-y^2}$ band intercepting the Fermi level, in particular, undergoes a $\sim 0.3$ eV reduction of its bandwidth, thereby eluding its apparent avoided crossing with the Ni~3$d_{z^2}$ band.
The O~2$p$ states, in contrast, are pulled down by $\sim$1.5 eV, thus increasing considerably the charge-transfer energy.
At the same time, the La~4$f$ states are shifted upwards by as much as 2 eV, so that they are removed from the vicinity of the Fermi level thus leading to a completely different optical absorption onset.

\paragraph{Computational methods.}
In all our calculations the crystal structure was that one of the space group 123 ($P4/mmm$ or $D_{4h}^1$) with the lattice parameters $a=3.96$ and $c=3.37$~\AA.
DFT-LDA calculations have been carried out with norm-conserving  pseudopotentials (PSP) on a plane-wave (PW) basis by the \textsc{abinit} code \cite{ABINIT}.
For the PSP-PW calculation we used a Troullier-Martins pseudopotential for O with six electrons in valence (2$s^2$ 2$p^4$), whereas for Ni we included 18 electrons in valence, that is, 3$d^8$ 4$s^2$ plus all semicore electrons 3$s^2$ 3$p^6$, and finally a Hartwigsen-Goedecker-Hutter pseudopotential \cite{HGH} for La with 11 electrons (5$d^1$ 6$s^2$ plus the semicore 5$s^2$ 5$p^6$) in valence.
While for DFT calculations the effect of including semicore electrons is negligible, this is not the case in $GW$ calculations for semicore electrons having a large spatial overlap with valence electrons.
The self-consistent DFT-LDA calculation was at convergence within 1~mHa in the total energy with a cutoff of 80~Ha, a $k$-point sampling of the Brillouin zone of $ 4 \times 4 \times 5$ shifted by $1/2 \times 1/2 \times 1/2$, and a Gaussian smearing of 0.01~Ha.
Pseudopotentials and DFT-LDA PSP-PW calculations have been validated with an all-electron full-potential linear-augmented plane-wave (FP-LAPW) calculation by the \textsc{wien2k} code \cite{WIEN2k}.
On top of the plane-wave DFT-LDA calculation we performed a one-shot $G_0W_0$ calculation using a Godby-Needs single plasmon-pole approximation for the screening by the \textsc{abinit} code.
A convergence of 0.1~eV on $GW$ quasiparticle energies has been achieved using a cutoff of 50~Ha on the wavefunctions and on the exchange part $\Sigma_x$ of the self-energy, 14~Ha for the size of dielectric matrices to take into account local-field effects and for the correlation part $\Sigma_c$ of the self-energy, 200 and 250 bands in the calculation of, respectively, the screening and the self-energy.
For the $GW$ calculation we used an unshifted $k$-point sampling of $4 \times 4 \times 4$ which includes all high-symmetry $k$ points.
$GW$ corrections are calculated in first-order perturbation theory with only diagonal matrix elements of the self-energy whose energy dependence is linearized and checked on nine frequencies.
The $GW$ Fermi level is recalculated on the $4 \times 4 \times 4$ grid at the end of the quasiparticle calculation.
Both DFT-LDA and $GW$ bands have been interpolated with maximally localized Wannier functions (MLWFs) by the \textsc{wannier90} code \cite{Wannier90}, considering separately the O~2$p$ band manifold below the Fermi energy and separated by a band gap from one side, and all the rest (Ni~3$d$, La~6$s$, La~5$d$, La~4$f$ Ni~4$s$ and Ni~4$p$) of valence electrons for the other side.

\begin{figure}[t]
\centering
\includegraphics[width=0.91\columnwidth]{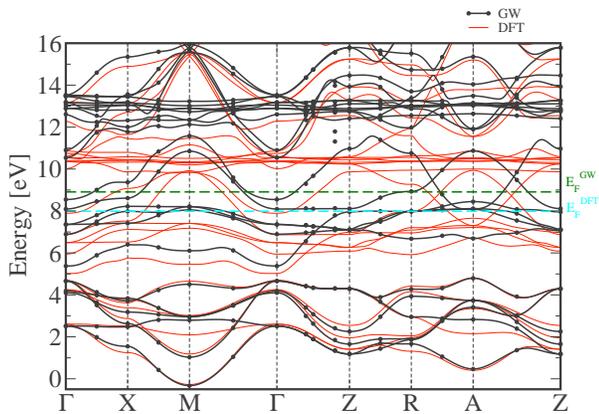}
    \caption{LaNiO$_2$ band plots as calculated in $GW$ (black dots) and interpolated by MLWF (black lines) with respect to DFT-LDA (red lines).
    In the left panel the horizontal cyan and green lines are the Fermi levels in DFT-LDA and $GW$, respectively.
    }
    \label{dftvsgw}
\end{figure}

\paragraph{Results.}
In Fig.~\ref{dftvsgw} we report the $G_0W_0$ band plot compared to DFT-LDA. 
In this figure the Fermi levels are not realigned and are indicated for both DFT and $GW$ as dashed lines, so that we can appreciate the effect of the $GW$ corrections to the zero-order DFT-LDA energies.
We can clearly see that the manifold of the lower six valence bands of O~2$p$ atomic character mainly (see, e.g., Ref.~[\onlinecite{bernardini19a}] for the atomic-orbital character of the bands), are practically unaffected by $GW$ corrections, apart from a flattening of an intermediate band at the $M$ and the $A$ points.
On the other hand, we observe important $GW$ corrections for the next manifold of the mainly Ni~3$d$ character bands closer to the Fermi level, and a rather large shift upwards of almost 3 eV of the seven La~4$f$ flat bands above the Fermi energy.

\begin{figure}[t]
\centering
\includegraphics[width=\columnwidth]{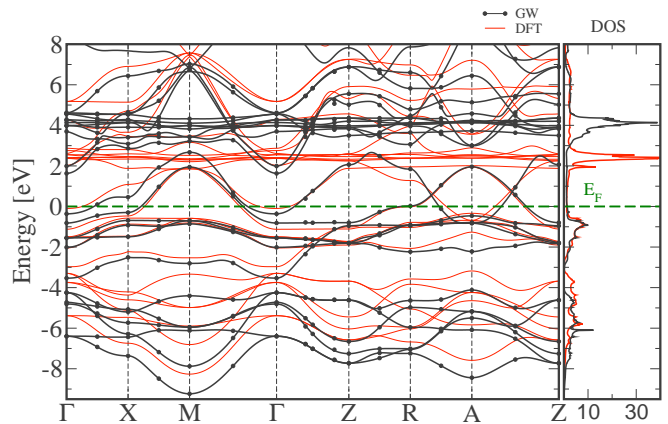}
    \caption{Same LaNiO$_2$ band plots as in Fig.~\ref{dftvsgw} but with both DFT and $GW$ Fermi levels realigned to zero.
    We also present the DFT and $GW$ density of states DOS.
    }
    \label{dftvsgwbnddos}
\end{figure}

\begin{figure*}[t!]
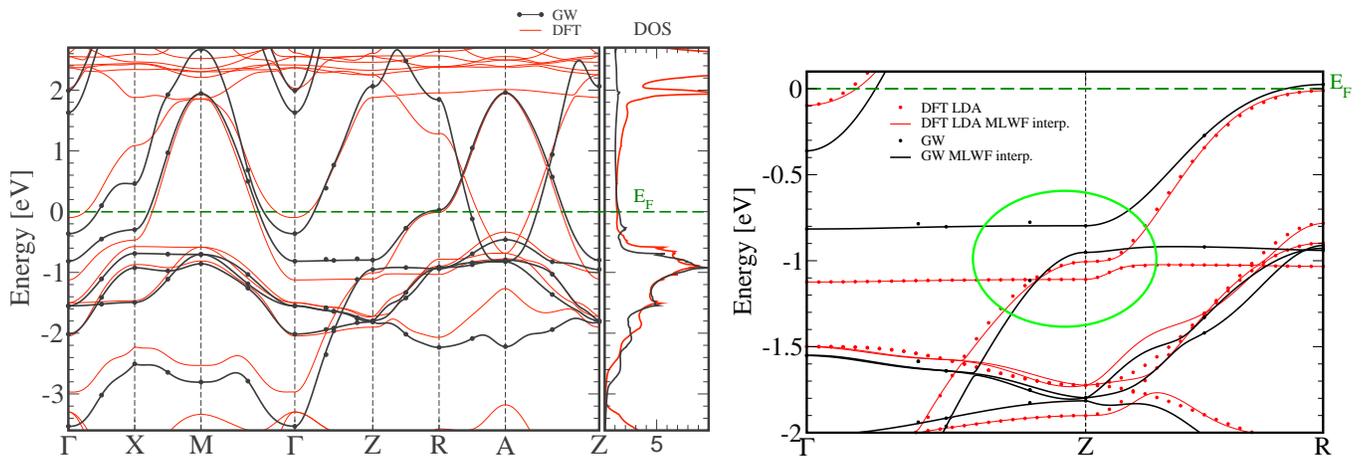

\includegraphics[width=.52\textwidth]{lanio2bndzoom}\hspace{.01\textwidth}
\includegraphics[width=.46\textwidth]{lanio2bndcross}
    \caption{LaNiO$_2$ band plots near the Fermi level as calculated by DFT-LDA (red) and $GW$ (black). The lines represent MLWF interpolations and the Fermi levels are all aligned to zero. Left: Entire dispersion of $d$-like bands around the Fermi level. Right: Zoom along the $\Gamma$-$Z$-$R$ path emphasizing the avoided crossing region. 
    }
    \label{dftvsgwzoom}
\end{figure*}

\begin{figure}[b]
    \includegraphics[width=0.45\columnwidth]{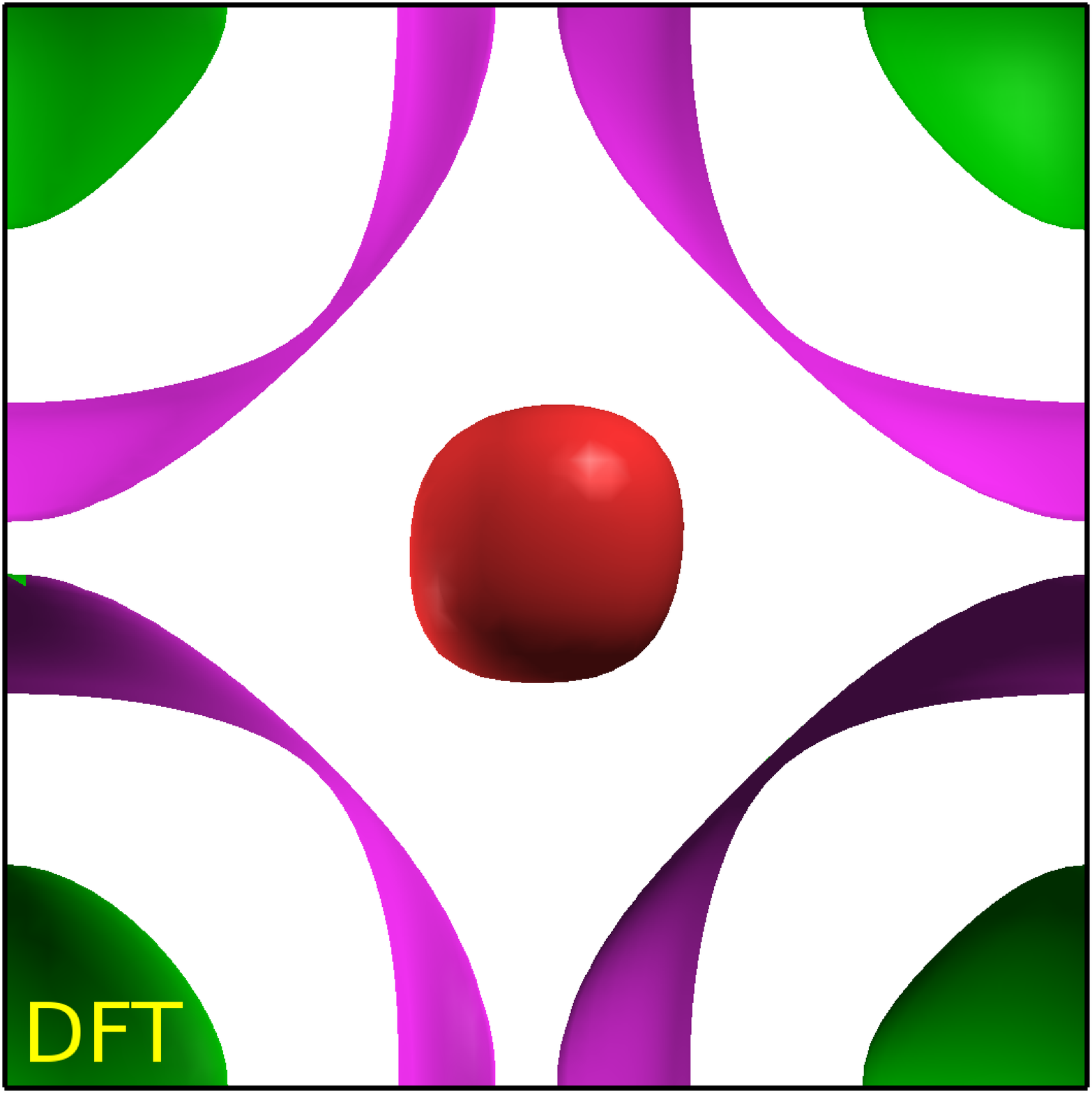}
    ~
    \includegraphics[width=0.45\columnwidth]{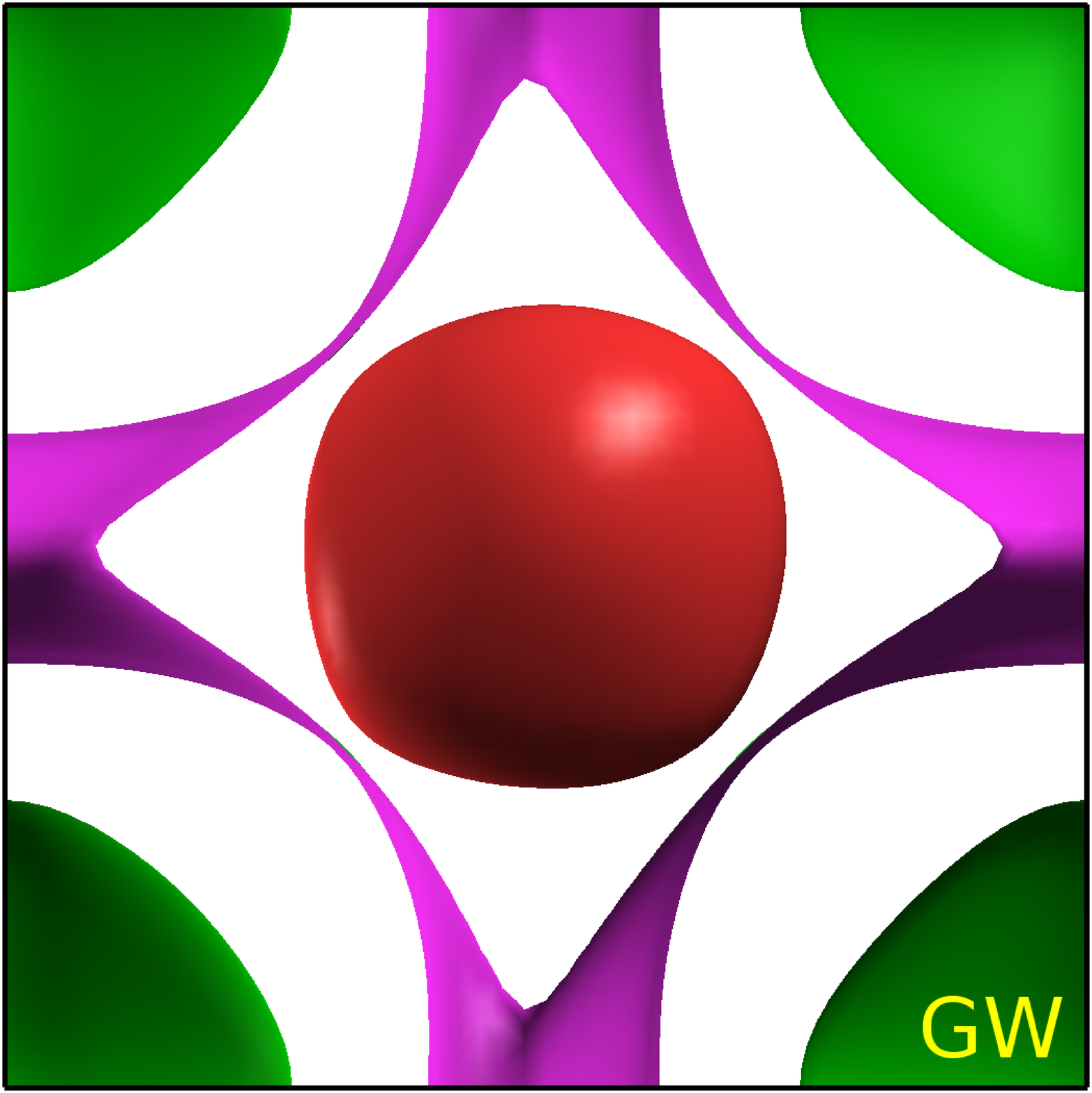}
    \caption{LaNiO$_2$ Fermi surfaces for DFT LDA (left) and $GW$ (right), as in a $z$ projection on an $xy$ plane bottom.
    }
    \label{fermisurface}
\end{figure}

However, when the Fermi energies are realigned as in Fig.~\ref{dftvsgwbnddos}, we surprisingly discover that $GW$ correlations do not induce qualitative modifications within a $\pm 2$~eV window around the Fermi level (see also Fig.~\ref{dftvsgwzoom} left).
In fact, the resulting Fermi-surface topology is exactly as in DFT-LDA (see Fig.~\ref{fermisurface}).
In both cases there is a large sheet ---almost dispersionless along $k_z$--- together with two electron pockets around the $\Gamma$ and $A$ points.
In this respect, the main $GW$ correction is an increased size of the electron pocket at $\Gamma$ as the corresponding band dips down to $-$0.35~eV below the Fermi level (see Fig.~\ref{dftvsgwzoom}).
This makes that pocket more ``resistant'' to hole doping and therefore can have a quantitative effect on the relative positions of the boundaries in the phase diagram of LaNiO$_2$.
At the same time, the bottom of this band remains just 0.1 eV above the top of the hole like band at $A$ that can also show up by doping (see Fig.~\ref{dftvsgwzoom}).
We note also that the large Fermi-surface sheet at $k_z = \pm \pi/c$ wraps around $A$ in DFT-LDA while it does around $R$ in $GW$.
A close inspection of the band plot reveals that this feature can be affected by a tiny shift of the Fermi level of just only 0.01~eV (see Fig.~\ref{dftvsgwzoom}). 
This degree of accuracy, however, is too high even for the DFT calculations.
In fact, our FP-LAPW and pseudopotential calculations show the same difference and, in any case, such a degree of accuracy also seems too high from the experimental point of view.

On the other hand, despite $GW$ corrections being negligible for them, the O~2$p$ bands are shifted down by $\sim$1~eV due to Fermi-level realignment only.
In $GW$ they appear further away from the Fermi level.
Also, the band gap between this O~2$p$ manifold and the topmost occupied bands opens further.
Consequently, many-body effects increase the corresponding charge-transfer energies.

A similarly large effect is observed for the La~4$f$ flat bands of localized electrons which have been suggested to be important for understanding some key properties of the superconducting nickelates \cite{ChoiPickett20}.
The La~4$f$ are pushed up at $\sim$4~eV above the Fermi level in $GW$.
This rising ``uncovers'' the La~5$d$ and 6$s$ bands, making them more important for the low-energy physics.
In fact, they are even downshifted in $GW$ compared to DFT (see, e.g., the states at $\Gamma$ at $\sim$2~eV above the Fermi level in Figs.~\ref{dftvsgw} and \ref{dftvsgwzoom}).
Thus, the La~6$s$ states become the first completely empty states above the Fermi level at $\Gamma$ and hence provide the first direct optical transitions according to $GW$.

In Fig.~\ref{dftvsgwzoom} we point out another probably significant modification induced into the electronic structure when introducing correlations by a $GW$ self-energy.
As we can see, the DFT-LDA band structure seems to display a band crossing along the $\Gamma$-$Z$ path at $\sim$1~eV below the Fermi level, immediately followed by an avoided crossing along the $Z$-$R$ path.
The $GW$ band plot, in contrast, does not display this feature.
As a result, the entire Fermi-surface sheet shown in pink in Fig.~\ref{fermisurface} comes from the same band with a strong Ni~3$d_{x^2-y^2}$ character everywhere. 
The most puzzling point is that also the DFT-LDA band plot presents in fact the same anti-crossing structure as in $GW$.
Contrary to appearances, there is no band crossing in DFT-LDA either.
Indeed, a band crossing in this region would imply that the large Fermi surface sheet drawn in pink in Fig.~\ref{fermisurface} would arise not from a single but from two different bands: the first crossing the Fermi level near $X$, and the second at $R$.
Nevertheless, we have checked that both in $GW$ but also in DFT-LDA there is only one large Fermi-surface sheet originating from only one single band.
We note that this feature is just only $-$0.8 eV from the Fermi energy and at $-$0.4 eV from the bottom of the electron pocket at $\Gamma$. 
Consequently, it might play an important role under the physical doping that possibly leads to superconductivity.

We tried to reproduce our $GW$ calculation using hybrid functionals, such as LDA0, PBE0, B3LYP, B3PW91, and also LDA+$U$.
All these hybrids keep the same LDA/PBE Fermi-surface topology, but none of them is able to reproduce the $GW$ shifts of the La~4$f$ and O~2$p$ states.
In LDA+$U$, in addition, the  differences with $GW$ are surprisingly very large on Ni~3$d$ states, too.
These technical details will be reported elsewhere.

Finally, we compare our \textit{ab initio} $GW$ electronic structure with recently reported dynamical mean-field theory (DMFT) calculations \cite{RyeeHan19,lechermann19,KarpMillis20}.
Note that $GW$ is an approximation to the exact solution, while DMFT relies on a Hubbard model (with the $U$ term on the Ni~3$d$ electrons only in the case of Ref.~[\onlinecite{RyeeHan19}], for example).
In both $GW$ and DMFT, the DFT-LDA Fermi-surface topology is preserved, with small differences regarding the size of the electron pocket at $\Gamma$ (a small stretch in $GW$, a tiny shrinkage in DMFT \cite{RyeeHan19,KarpMillis20}).
The bandwidth of the Ni~3$d_{x^2-y^2}$ band is also reduced in both cases, although this reduction can be far more pronounced in the DMFT case \cite{RyeeHan19,KarpMillis20} due to the neglect of the nonlocal part of the self-energy \cite{MiyakeAryasetiawan13,TomczakBiermann14}.
When it comes to the quasiparticle weight $Z$ associated with this band, $GW$ yields an almost constant $Z^{GW}=0.70 \pm 0.02$ all along the Brillouin zone, whereas in Ref.~\onlinecite{RyeeHan19}] the reported $Z$ displays a noticeable variation and is considerably reduced at $\Gamma$ and along $\Gamma$-$Z$-$R$.
$Z$ is also rather constant in Ref.~[\onlinecite{KarpMillis20}], but here a lower/upper Hubbard band splitting is clearly visible, unlike in Ref.~[\onlinecite{RyeeHan19}] and in our work.
The $Z$ of the additional La~5$d$ band crossing the Fermi level is similar in $GW$ and DMFT. Specifically, we find $Z^{GW}=0.79 \pm 0.02$ everywhere but at $A$ where it goes down to $Z^{GW}=0.71$.
The main differences between $GW$ and DMFT are observed in the rest of the band plot. The other Ni~3$d$ bands, in particular, are sensibly different and the La~4$f$ states remain unshifted in DMFT. More importantly, $GW$ and DMFT shift the O~2$p$ states in opposite directions, which makes a qualitative difference on the corresponding changes in the charge-transfer energies.

The relatively small changes introduced by $GW$ many-body effects near the Fermi level make possible a tight-binding fit of the dominant $d_{x^2-y^2}$ band as in Ref.~[\onlinecite{botana19}] (see Fig.~\ref{tb} in the Appendix). 
This allows us to define the ratio $t'/t = (|t_2|+|t_3|)/|t_1|$ between the longer-range hoppings to the nearest-neighbor hopping.
It has been argued that a larger $t'/t$ ratio correlates to a higher $T_c$ \cite{pavarini-prl01}.
The parameters of the fit are summarized in Table \ref{t:tb}. 
As we can see, many-body effects slightly renormalize the $t'/t$ ratio from $0.37$ \cite{botana19} to $0.33$. 
The fit can be extended to include the second band crossing the Fermi level as described in the Appendix.
This further reduces the $t'/t$ ratio to $0.24$.

\begin{table}[tb!]
\begin{tabular*}{\columnwidth}{@{\extracolsep{\fill}} r p{1em} l}
\hline \hline 
$t_n$ (meV) && $f_n({\mathbf k})$ \\
\hline
374&& 1 \\
$-344$ && $2 [\cos(k_x a) +  \cos(k_y a)]$ \\
81 && $4 \cos(k_x a)   \cos(k_y a)$ \\
$-33$ && $2 [\cos(2 k_x a) +  \cos(2 k_y a)]$\\
$-172$ && ${1\over 4} [\cos(k_x a) -  \cos(k_y a)]^2 \cos(k_z c)$ \\
$-71$ && ${1\over 4} [\cos(k_x a) -  \cos(k_y a)]^2 \cos(2 k_z c)$ \\
\hline \hline
\end{tabular*}
\caption{Tight-binding fit of the $GW$-corrected Ni~3$d_{x^2-y^2}$ band at the Fermi energy with $\varepsilon({\mathbf k}) = \sum_n t_n f_n ({\mathbf k})$.}
\label{t:tb}
\end{table}

\paragraph{Conclusions.}
We have calculated the \textit{ab initio} $GW$ correlated electronic structure of LaNiO$_2$.
$GW$ many-body effects do not affect the DFT-LDA Fermi surface topology and only slightly renormalize the size of the electron pockets.
La~4$f$ states undergo a 2 eV shift that place them faraway from the Fermi level.
O~2$p$ states downshift so as to increase the ionic charge-transfer character.
We observe also a missed band crossing just below the Fermi level and at a distance which can be relevant at the hole-doping levels at which superconductivity occurs.
To date, angle-resolved photoemission (ARPES) data on LaNiO$_2$ is lacking.
Thus our $GW$ calculation provides a genuine prediction of the electronic structure of LaNiO$_2$.
Future ARPES experiments will represent a check of the $GW$ approximation validity.

\paragraph{Note added.}
Recently, we became aware of a $GW$ calculation on LaNiO$_2$ (see Appendix~C and Fig.~9 of Ref.~[\onlinecite{PhysRevB.101.075107}]) done without La~4$f$ electrons.
The latter is the reason for major differences between the two calculations.

\paragraph{Acknowledgments.}
F.B. acknowledges financial support by the Visiting Scientist Program of the Centre de Physique Theorique Grenoble-Alpes (CPTGA). 

\appendix
\paragraph{Appendix.}

In this appendix we provide a validation of our norm-conserving pseudopotentials and of our PSP-PW calculation with respect to an all-electron full-potential linear-augmented plane-wave (FP-LAPW) calculation by the code \textsc{wien2k} \cite{WIEN2k}.
We also discuss the validity of our MLWF set and interpolation. and provide two-band tight-binding fit of the main $GW$-corrected bands at the Fermi level of LaNiO$_2$.

For the FP-LAPW calculation we set muffin-tin radii to 2.5, 2.1 and 1.62~bohr for La, Ni and O respectively, a cut off of $R_\mathrm{MT}K_\mathrm{max} = 7.0$, and a Brillouin zone sampling of $11 \times 11 \times 14$ for the self-consistent calculation of the density.

\begin{figure}[b]
    \centering
    \includegraphics[width=\columnwidth]{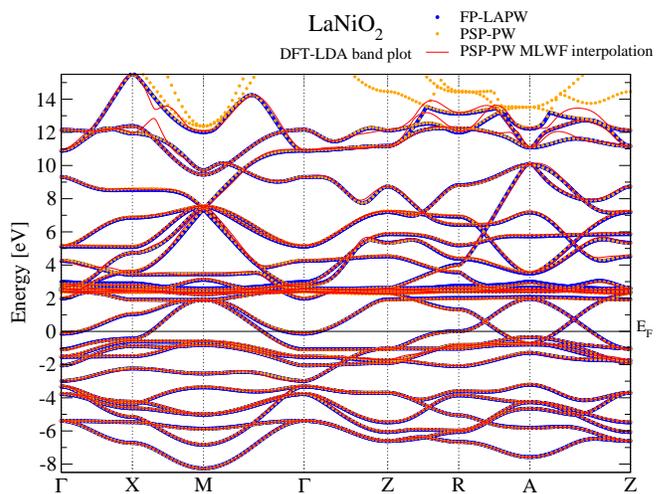}
    \caption{LaNiO$_2$ DFT-LDA band plot calculated in full-potential linear-augmented plane-waves (FP-LAPW, blue dots) by the \textsc{wien2k} code, as compared to norm-conserving pseudo-potential plane-waves bands (PSP-PW, orange dots) by the \textsc{abinit} code.
    The PSP-PW bands have been interpolated by maximally-localized Wannier functions (MLWF, red lines) by the \textsc{wannier90} code.
    The Fermi level is set to zero.
    }
    \label{dftabinitwien2k}
\end{figure}

In Fig.~\ref{dftabinitwien2k} we compare the DFT-LDA band plots calculated using the FP-LAPW and the PSP-PW methods.
Although the calculations have been carried out using very different methods and convergence parameters, we found a very good agreement between them.
This is an important validation of the pseudopotentials as well as of the convergence parameters used in the plane-waves calculation which constitutes the starting point of our $GW$ calculation.
Our DFT band plot is also in very good agreement with previous calculations \cite{botana19,sakakibara19}.

\begin{figure*}[t]
    \includegraphics[width=0.45\textwidth]{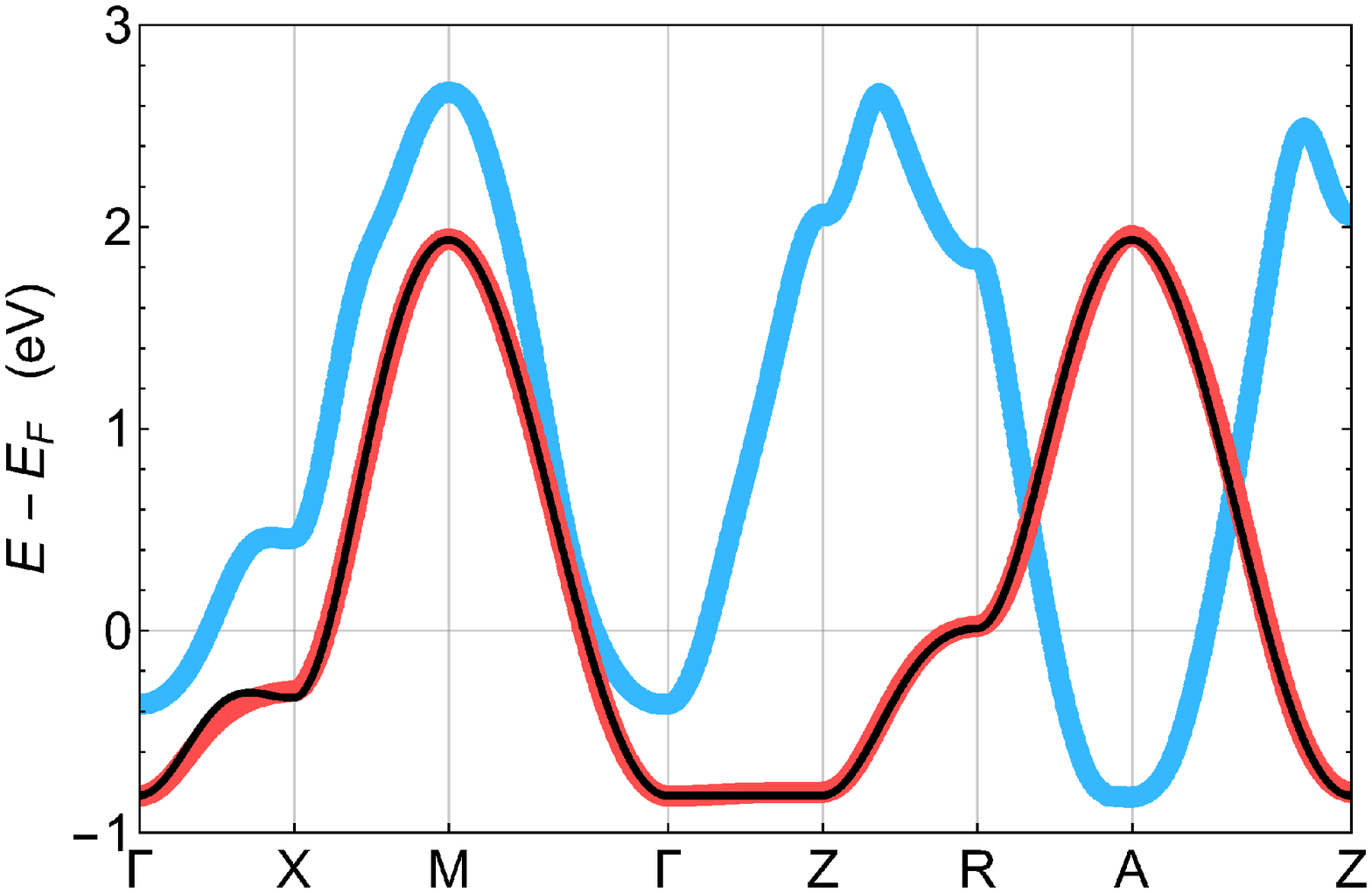}
    \hspace{.025\columnwidth}
    \includegraphics[width=0.45\textwidth]{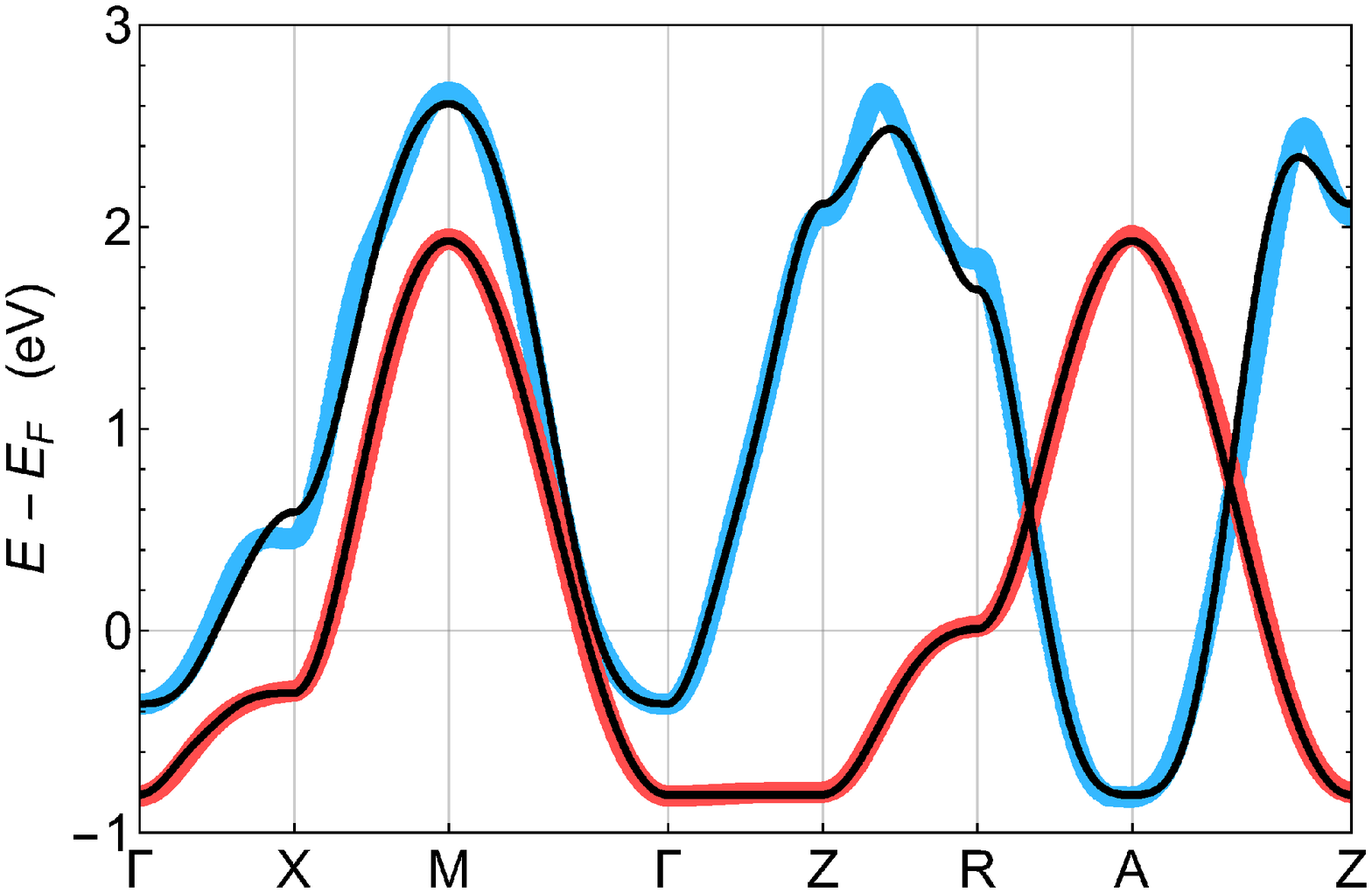}
    \caption{Tight-binding fits (black) of the main $GW$ bands at the Fermi energy (red and blue) of LaNiO$_2$. }.
    \label{tb}
\end{figure*}

In the same figure we report also the norm-conserving PSP-PW bands as interpolated using maximally-localized Wannier functions (MLWF).
To build MLWF we used an unshifted $4 \times 4 \times 4$ $k$-point grid, different from the grid we used to calculate the self-consistent DFT-LDA density.
One can see that the interpolation is satisfactory and starts to deviate on the highest energies, more than 10~eV bands above the Fermi level.
There are however some fine details, not visible at the scale of Fig.~\ref{dftabinitwien2k}, such as some band crossings, which are missed by the MLWF interpolation.
For this reason, whenever a critical detail / band crossing was concerned, we referred to the real direct band plot calculations instead of the MLWF interpolation.

Since in $G_0W_0$ the wavefunctions are kept at the DFT-LDA level and not updated, the MLWF are not updated neither.
So the quality of the $GW$ band interpolation should be at the same level of DFT-LDA, and this can be checked in Fig.~\ref{dftvsgw}.
Like DFT-LDA, $GW$ energies are also reproduced remarkably well.
However, there are two $GW$ energies along the $\Gamma$-$Z$ direction not reproduced by the interpolated bands (at 11.3 eV and 11.8 eV).
This $k$ point does not belong to the unshifted $4 \times 4 \times 4$ grid which was used to calculate the MLWF, and belongs to a grid which was introduced to check the MLWF and effective band crossings. 
Thus $GW$ bands might look differently from what represented by the MLWF interpolation in that $\Gamma$-$Z$ $E=[11.3,11.8]$ eV region.
We recommend future ARPES experiments to keep into account this issue when comparing to our $GW$ bands.

The tight-binding fit of the main band described in the main text is shown in Fig.~\ref{tb} (left panel). 
In addition, we extended the fit to the second band that crosses the Fermi level by supplementing the tight-binding Hamiltonian with the matrix elements
\begin{widetext}
\begin{align}
\varepsilon_{\rm La}= \; & t_{000} 
+ 2 t_{001} \cos(k_z ) + 2 t_{002} \cos(2 k_z ) +2 t_{003} \cos(3 k_z )
\nonumber \\
&+[2 t_{100} + 4 t_{101} \cos(k_z) +  4 t_{102} \cos(2k_z) ] [\cos(k_x) + \cos(k_y)]
\nonumber \\
&+[4 t_{110} + 8 t_{111} \cos(k_z)   
+ 8 t_{113} \cos(3k_z)] \cos(k_x) \cos(k_y)
\nonumber \\
&+[4 t_{210} + 8 t_{211} \cos(k_z)] [\cos(2 k_x)\cos(k_y) + \cos(k_x)\cos(2k_y)],
\\
\varepsilon_\text{La-Ni}  =\; & 8 t ^\text{La-Ni}  [\cos({3k_x / 2})\cos({k_y/2})-\cos({k_x / 2})\cos({3k_y / 2})]\cos({k_z / 2}) .
\end{align}
\end{widetext}
The parameters for this fit are shown in Table~\ref{tbparam}, and the fit is displayed in Fig. \ref{tb} (right panel).

\begin{table}[h!]
\begin{tabular}{|l p{1em} r | l p{1em} r | p{1em}  l |}
\hline
$_n$ && $t_n$ & $_{ijk}$ && $t_{ijk}$ && $t^\text{La-Ni}$ \\
\hline
$_0$ && 420    & $_{000}$ && $1299$ && $-0.016$\\
$_1$ && $-343$ & $_{001}$ && $-135$ &&\\
$_2$ && 61     & $_{002}$ && $-139$ &&\\
$_3$ && $-24$  & $_{003}$ && 56 &&\\
$_4$ && $-159$ & $_{100}$ && $-43$ &&\\
$_5$ && $-224$ & $_{101}$ && $-232$ &&\\
&&             & $_{102}$ && 43 &&\\
&&             & $_{110}$ && $-32$ &&\\
&&             & $_{111}$ && 72 &&\\
&&             & $_{113}$ && $-23$&&\\
&&             & $_{210}$ && $-23$&&\\
&&             & $_{211}$ && 24 && \\
\hline
\end{tabular}
\caption{Parameters of the two-band tight-binding model (units are meV).}
\label{tbparam}
\end{table}

\bibliography{nickelates}

\end{document}